\documentclass[aps,pra,reprint,nofootinbib,superscriptaddress,twocolumn,showpacs,showkeys,longbibliography,amsmath,amssymb]{revtex4-1}
\usepackage{graphicx}
\usepackage{dcolumn}
\usepackage{bm}
\usepackage{braket}
\usepackage{subfigure}
\usepackage[colorlinks,bookmarks=false,citecolor=blue,linkcolor=red,urlcolor=blue]{hyperref}
\usepackage[english]{babel}
\usepackage{changes}
\usepackage{CJK}

\begin{document}

\title{Effect of anisotropic spin-orbit coupling on condensation and superfluidity of a two dimensional Fermi gases}

\author{Kezhao Zhou}
\email{kezhaozhou@gmail.com}

\affiliation{Department of Physics, College of Science, Hunan University of Technology, Zhuzhou 412007, China}

\date{\today}

\begin{abstract}
We investigated the ground state properties of a two dimensional Fermi superfluid with an anisotropic spin-orbit coupling (SOC) using path-integral field theoretical method. Within the framework of mean-field theory, we obtained the condensed fraction including contributions from both singlet and triple pairing fields. We found that for small interaction parameters and large anisotropic parameters, the total condensed fraction changes non-monotonically when increasing the strength of SOC and has a global maximum. But this feature disappears with decreasing the anisotropic parameter and increasing the interaction parameter. However, condensed fraction always decrease with increasing the anisotropic parameters. Because of the anisotropy of the SOC, the superfluid fraction becomes a tensor. We obtained the superfluid fraction tensor by deriving the effective action of the phase field of the order parameter. Our numerical results show that for small interaction parameters and large anisotropic parameters, superfluid fraction of the $x$ component $\rho_{x}$ has a minimum as a function of the SOC strength. And this minimum of $\rho_{x}$ disappears when decreasing the anisotropic parameters. In the strong interaction regime, $\rho_{x}$ always decreases with increasing the strength of SOC. While for the $y$ component of the superfluid fraction $\rho_{y}$, no matter how large the interaction parameters and anisotropic parameters are, it always has a minimum as a function of the SOC strength. As a function of the anisotropic parameter, for strong SOC strength, $\rho_{x}<\rho_{y}$ with $\rho_{x}$ having a minimum. For small SOC parameters $\rho_{x}>\rho_{y}$ with $\rho_{y}$ developing a minimum only in the weak interaction limit.

\end{abstract}

\maketitle

\section{Introduction.}

Spin-orbit coupling (SOC) plays a essential role in realizing many novel phenomena such as topological superconductors and insulators \cite{nayak, qi, topobook,volovik}, Floquet topological phases \cite{lindner,rudner,xus}, nontrivial superconductors \cite{bergeret, luyang} and so on. Realization of SOC in ultra-cold atomic system \cite{lin, williams, pengjun, pan} using Raman couplings has attracted lots of interest in various physics community. Because of advances in ultra-cold atoms experimental techniques, Fermi gases with SOC provides a unique and important playground to investigate various novel phases and topological phase transitions. For example, using Feshbach resonances \cite{hara, bourdel}, one can tune interactions between atoms from weakly interacting regime to strong interaction regime, driving the system from weakly interacting BCS superfluid to strongly interacting BEC regime (BCS-BEC crossover)\cite{giorgini,bcs-bec-chapter}. Furthermore, optical lattice trapping potential makes this system a perfect platform to mimic solid state systems and related phenomena \cite{lewenstein}. In experiments, the topological band structure has been observed by combinations of optical lattice and SOC\cite{meng, cooper}. Other experimental techniques, such as spectroscopy \cite{cheuk,spec-jin,spectroscopy-torma}, dipole interactions \cite{dipole-fermi,dipole-review}, reduced dimensions \cite{lowd-review}, dynamical quench \cite{noneq-review}, open quantum systems \cite{open} and so on, are also used to detect various phenomena related with Fermi pairing and superfluid \cite{dalibard,goldman}.

In ultra-cold atomic systems, one can create any kind of SOC in principle, especially the Rashba \cite{rashba, gorkov} and Dresselhaus \cite{gds} SOC. Current experimental set-up can produce SOC with arbitrary combination of these two types of SOC \cite{pengjun, pan}, therefore create an anisotropic SOC. Along this line, Many theoretical investigations have been performed to study effects of SOC on various superfluid properties \cite{xiang, sala, lee, wu, setti, zhair, zhangjing, sun, lianyi2, zhangk, zinner, shenoy, vjs, iskin, cuij, zheng, lianyi1, zhaihui, zhangyicai, su, koinov, zengqiang,iskin-new}. For the balanced case with equal particle numbers of different internal states, SOC can produce a novel bound-state called Rashbons and induce a crossover from weakly correlated BCS to strongly interacting BEC regime even for very weak particle-particle interaction \cite{shenoy,vjs}. And this new bound states have many important implications on various thermodynamic properties of the system. Especially, the opposite effect of SOC on condensation and superfluidity has been discussed in ~\cite{zhou}. Furthermore, combined effect of SOC and Zeeman field can host a non-trivial topological order \cite{ghosh, sau, carlos, liuxj, cao, devreesej, luo, iskins, lianyi3, xuyong, gong}. Besides, the presence of Zeeman field can create a novel FFLO phase which attracts lot of interests in superconductors and cold atomic system \cite{kin}. For Fermi gases with SOC, a new type topological FFLO state has also been investigated extensively \cite{zheng,caoye, zhang,dongli}. Effects of anisotropic SOC on the ground state properties have also been discussed in \cite{sala}. And in \cite{devreesej}, effects of anisotropic SOC on BKT \cite{bkt} transitions and collective sound velocity have been investigated. Furthermore, \cite{iskin-new} provides an intrinsic link between the non-monotonic behavior of the superfluid density and the quantum geometry of the helicity bands.

In this paper, we performed a detailed research on the effects of an anisotropic SOC on the condensation and superfluidity of a two-dimensional (2D) superfluid system within the framework of mean-field theory using path-integral formalism. The coupled number and gap equations are numerically solved to get the chemical potentials and gap parameters. With the obtained chemical potentials and gap parameters, we calculated the condensed and superfluid fraction as functions of the interaction parameter, SOC strength and anisotropic parameters. For the condensed fraction, we consider contributions from both the singlet and triplet pairing fields. As a function of the SOC parameters, the condensed fraction behaves non-monotonically for specific interaction, SOC strength and anisotropic parameters. In order to obtain the superfluid fraction, we expand the partition function to the quadratic order of the phase of the order parameter from which we read off the superfluid fraction. The superfluid fraction is a tensor because of the anisotropic SOC considered in this paper. And our numerical results show that different components of the superfluid tensor behave differently as functions of the SOC and anisotropic parameters.

\section{Formalism.}

The system we consider here is a 2D ultra-cold Fermi atoms or electrons interacting attractively with a contact interaction. We also consider an anisotropic SOC which can be written as an arbitrary combination of Rashba and Dresselhaus type of SOC. In the path-integral formalism, the system can be described by the finite temperature grand-partition function $Z=\int d[\bar{
\varphi}_{\sigma },\varphi _{\sigma }]\exp \left( -S[\bar{\varphi}_{\sigma },\varphi
_{\sigma }]\right) $ ($\hbar =k_{B}=1$ through out this paper) with the action $S[\bar{
\varphi}_{\sigma },\varphi _{\sigma }]$ being given by
$S[\bar{\varphi}_{\sigma},\varphi_{\sigma }]=\int_{0}^{\beta }d\tau \int d^{2}\mathbf{r}\sum_{\sigma }[\bar{
\varphi}_{\sigma }\partial _{\tau }\varphi _{\sigma }+\mathcal{H}_{0}+\mathcal{H}
_{I}]$ with $\beta =1/T$ , $\sigma =\uparrow ,\downarrow $ denoting the two different internal states of the atoms or z component eigen-states of the spin operator for electrons and $
\bar{\varphi}_{\sigma }, \varphi _{\sigma }$ being the Grassmann fields. The single-particle Hamiltonian density is $
\mathcal{H}(\bar{\psi},\psi )\mathcal{=}\bar{\psi}\left( \hat{\xi}_{\mathbf{p
}}+\mathcal{H}_{soc}\right) \psi $ where the kinetic operator
$\hat{\xi}_{\mathbf{p}}=\mathbf{\hat{p}}^{2}/(2m)-\mu $ with $\mu $ being
the chemical potential fixed by the total particle number, the spinor field reads: $\psi \left( \mathbf{r}\right) =
\left[ \varphi _{\uparrow }\left( \mathbf{r}\right) ,\varphi _{\downarrow
}\left( \mathbf{r}\right) \right] ^{T}$ and the SOC term can be written as:
\begin{equation}
\mathcal{H}_{soc}=\lambda _{R}\left( \sigma _{x}p_{y}-\sigma _{y}p_{x}\right) +\lambda
_{D}\left( \sigma _{x}p_{y}+\sigma _{y}p_{x}\right)   \label{soc}
\end{equation}
where $\lambda _{R}$ and $\lambda _{D}$ denote the Rashba and Dresselhaus SOC
parameters respectively and $\sigma _{i=x,y,z}$ are the Pauli matrices. In order to show the anisotropic character transparently, the SOC term can be re-written as: $\mathcal{H}_{soc}=\lambda _{y}\sigma _{x}p_{y}+\lambda
_{x}\sigma _{y}p_{x}$ with $\lambda _{y}=\lambda _{D}+\lambda _{R}$ and $\lambda _{x}=\lambda _{D}-\lambda _{R}$. From this definition, we can see that the system is isotropic when $\lambda _{D}=0$ or $\lambda _{R}=0$ and anisotropic for equal Rashba and Dresselhaus (ERD) SOC: $\lambda_{D}=\lambda _{R}$. For convenience, we define the anisotropic parameter as:
\begin{equation}
\eta =\frac{\lambda _{D}}{\lambda _{R}}
\end{equation}
Without loss of generality, when $\eta $ increases from $0$ to $1$, the system evolves from isotropic
Rashba case to anisotropic case with ERD SOC. We denote the SOC strength by:

\begin{equation}
\lambda = \sqrt{\lambda_{D}^2+\lambda_{R}^2}
\end{equation}

Finally, the interaction between spin-up and spin-down component can be simplified by a contact interaction model:
\begin{equation}
\mathcal{H}_{I}= -g\int d^{2}\mathbf{r}\varphi _{\uparrow }^{\dagger
}\left( \mathbf{r}\right) \varphi _{\downarrow }^{\dagger }\left( \mathbf{r}
\right) \varphi _{\downarrow }\left( \mathbf{r}\right) \varphi _{\uparrow
}\left( \mathbf{r}\right)
\end{equation}
where $g>0$ is the contact interaction parameter.

Within path integral methods, the pairing order parameter can be conveniently introduced by using the Hubbard-Stratonovich transformation \cite{altland} to decompose the four-body interaction term $\mathcal{H}_{I}$ by introducing a pairing field $\Delta\left( \mathbf{r},\tau \right) $. After integrating out the fermionic fields, we obtain the effective action of the pairing field as $S_{eff}\left[ \bar{\Delta},\Delta \right] =-\int_{0}^{\beta }d\tau \int d^{d}\mathbf{r}\left\vert\Delta \left( \mathbf{r},\tau \right) \right\vert ^{2}/g-1/2Tr\ln \left[\mathcal{G}_{\mathbf{r},\tau }^{-1}\right] $ with the inverse Greens' function $\mathcal{G}_{\mathbf{r},\tau }^{-1}$ being:
\begin{equation}
\mathcal{G}_{\mathbf{r},\tau }^{-1}=\left[
\begin{array}{cccc}
\partial _{\tau }+\hat{\xi}_{\mathbf{p}} & \hat{\gamma}_{_{\mathbf{p}}} & 0
& \Delta \\
\hat{\gamma}_{_{\mathbf{p}}}^{\ast } & \partial _{\tau }+\hat{\xi}_{\mathbf{p}} & -\Delta & 0 \\
0 & -\bar{\Delta} & \partial _{\tau }-\hat{\xi}_{\mathbf{p}} & \hat{\gamma}_{_{\mathbf{p}}}^{\ast } \\
\bar{\Delta} & 0 & \hat{\gamma}_{_{\mathbf{p}}} & \partial _{\tau }-\hat{\xi}_{\mathbf{p}}
\end{array}\right]   \label{inversepropagator}
\end{equation}
with $\hat{\gamma}_{_{\mathbf{p}}}=\lambda_{y}\hat{p}_{y}+i\lambda_{x}\hat{p}_{x} $.

At mean-field level, the pairing field can be chosen as a real constant parameter $\Delta \left( \mathbf{r},\tau \right) =\Delta _{0}$
which is referred to as the gap parameter. And the effective pairing action
becomes $S_{eff}\left[ \bar{\Delta},\Delta \right] =-\beta V\Delta
_{0}^{2}/g-1/2\sum_{\mathbf{p},i\omega _{n}}\ln \left[ \det \mathcal{G}_{\mathbf{p},i\omega _{n}}^{-1}\right] $ where $\mathcal{G}_{\mathbf{p},i\omega _{n}}^{-1}$ is the fourier transformation of of Eq. ({\ref{inversepropagator}}) in the momentum-frequency domain, $V$ is the areal size of the system and $\omega _{n}=\left( 2n+1\right) \pi/\beta $ are the Fermi Matsubara frequencies. From $\det \mathcal{G}_{\mathbf{p},E}^{-1}=0$, the quasi-particle excitation spectrum can be obtained as $E_{
\mathbf{p},\pm }=\sqrt{\left( \xi _{\mathbf{p}} \pm \left\vert \gamma_{\mathbf{p}}\right\vert \right)^{2}+\Delta _{0}^{2}}$ and $E_{\mathbf{p},\pm }^{\prime }=-E_{\mathbf{p},\pm }$ where $\xi _{\mathbf{p}}=\epsilon _{\mathbf{p}}-\mu $ with $\epsilon _{\mathbf{p}}=\mathbf{p}^{2}/2m$ and $\gamma_{_{\mathbf{p}}}=\lambda_{y}p_{y}+i\lambda_{x}p_{x} $. The mean-field thermodynamic potential can be obtained using $\Omega =-1/\beta \ln Z$ and we have: $\Omega _{0}=-V\Delta _{0}^{2}/g+1/2\sum_{\mathbf{p},\delta }\left( \xi _{\mathbf{p}}
E_{\mathbf{p,}\delta }\right)
-1/\beta \sum_{\mathbf{p},\delta =\pm }\ln \left( 1+e^{-\beta E_{\mathbf{p},\delta }}\right)$. By variation of the thermodynamic potential with respect to the chemical potential and order parameter, we can easily obtain the mean-field gap and number equations:
\begin{equation}
\frac{1}{g}=-\frac{1}{V}\sum_{\mathbf{p,}\delta =\pm }\frac{\tanh \left(
\frac{\beta E_{\mathbf{p,}\delta }}{2}\right) }{4E_{\mathbf{p,}\delta }},
\label{gap}
\end{equation}
\begin{equation}
n=\frac{1}{2V}\sum_{\mathbf{p,}\delta =\pm }\left[ 1-\frac{\left( \xi _{
\mathbf{p}}+\delta \left\vert \gamma _{\mathbf{p}}\right\vert \right) \tanh
\left( \frac{\beta E_{\mathbf{p,}\delta }}{2}\right) }{E_{\mathbf{p,}\delta }
}\right]  \label{num}
\end{equation}
As usual, divergence of the integral over momenta in Eq. ({\ref{gap}}) is removed by
replacing contact interaction parameter $g$ by binding energy $E_{b}$
through $V/g=\sum_{\mathbf{p}}1/\left( 2\epsilon _{\mathbf{p}}+E_{b}\right) $.

For anisotropic Rashba SOC, Eq. ({\ref{gap}}) and Eq. ({\ref{num}}) are widely used to study the ground state and finite temperature properties of this novel system.
It was first shown by Gor'kov and Rashba \cite{rashba,gorkov} that, in the presence of a
weak SOC, a 2D superconductor supports both singlet and triplet pairing
fields. In ultra-cold atomic systems, this non-trivial physics was investigated in \cite{shenoy,vjs} and proposal for detecting this anisotropic superfluidity was given in \cite{huhui} through measurement of the momentum distribution and single-particle spectral function. On the other hand, SOC significantly enhances the pairing
phenomena as was shown by the exact two-body solutions \cite{shenoy,vjs} and many-body mean-field calculations
\cite{hanli}. The system can evolve from a BCS to a
BEC state driven by SOC even for very weak interactions. Various other properties has been investigated in detail. For anisotropic SOC, however, there are not so many publications concerning the ground state properties. In \cite{devreesej}, they investigated the effect of anisotropic SOC on the BKT transition and collective sound velocity for a 2D Fermi gases. In this paper, we focus on the effect of anisotropic SOC on condensation and superfluidity from a different point of view presented in \cite{iskin-new}.

\section{Condensed density}
For Fermi pairing and condensation, according to the concept of off-diagonal-long-range-order, the condensed density is generally
defined as \cite{yang,leggett}: $n_{c}=1/V\sum_{\mathbf{p,}ss^{\prime}}\left\vert \left\langle
\bar{\varphi}_{\mathbf{p},s}\varphi_{-\mathbf{p},s^{\prime }}\right\rangle\right\vert ^{2}$. For the system considered in this paper, the attractive interaction supports a singlet-pairing
field while SOC hybridizes spin degrees of freedom and induces triplet pairing simultaneously.
Within mean-field theory, spin-singlet and -triplet pairing
fields are given by
\cite{huhui}:
\begin{equation}
\left\langle \bar{\varphi}_{\mathbf{p,}\uparrow }\varphi_{-\mathbf{p,}\downarrow }\right\rangle =\Delta _{0}\sum_{\delta }\tanh
\left( \beta E_{\mathbf{p},\delta }/2\right) /\left(4E_{\mathbf{p},\delta }\right)
\end{equation}
\begin{equation}
\left\langle \bar{\varphi}_{\mathbf{p,}\uparrow }\varphi_{-\mathbf{p,}\uparrow }\right\rangle =-\Delta _{0}\left( \gamma
_{\mathbf{p}}/\left\vert \gamma _{\mathbf{p}}\right\vert \right)\sum_{\delta }\delta \tanh \left( \beta E_{\mathbf{p},\delta
}/2\right) /\left( 4E_{\mathbf{p},\delta }\right)
\end{equation}
respectively.
The spin-singlet contribution to the condensed fraction was first
discussed in \cite{sala} where it was shown to behave
non-monotonically with a minimum as a function of SOC strength for
weak enough interaction parameter. In our previous investigations \cite{zhou}, we include both singlet and triplet contributions to the condensed density and obtain:
\begin{equation}
n_{c}=\frac{\Delta _{0}^{2}}{4}\frac{1}{V}\sum_{\mathbf{p,\delta }}\frac{
\tanh ^{2}\left( \frac{\beta E_{\mathbf{p},\delta }}{2}\right) }{E_{\mathbf{p
},\delta }^{2}}  \label{condensed}
\end{equation}

At zero temperature, repulsive interactions between Fermi pairs (Bosons)
result in depletion of the condensate which is a familiar phenomenon for
interacting BEC systems. Therefore the condensed fraction (condensed density divided by the total density $n$) is always less than $1$.

\section{Superfluid density}

Unlike the condensed density, superfluid density is a dynamical properties of the system and a tensor in general. In Landau's theory of superfluidity, the normal
mass of the system can be obtained through the calculation of the
total momentum carried by excitations when the system is enforced in
a uniform flow with velocity \cite{landau-st-2} $\mathbf{v}_{s}$
\begin{equation}
\mathbf{P}=\sum_{\mathbf{p,}\sigma }\mathbf{p}f\left( E_{\mathbf{p,}\sigma }-\mathbf{p\cdot v}_{s}\right)   \label{landau}
\end{equation}
where $\sigma $ is a conserved quantum number which is spin in the
absence of SOC, $f\left( x\right) =1/\left( e^{\beta x}\pm 1\right)$ is the Fermi/Bose distribution function depending on the nature of
the excitations, and $E_{\mathbf{p,}\sigma }-\mathbf{p\cdot v}_{s}$ is the excitation spectrum for moving systems. At zero temperature, no excitations are created at very small $\mathbf{v}_{s}$ and the superfluid density coincides with the total density.

However, the situation is dramatically changed in the presence of SOC
where Galilean transformation is violated. As pointed out in \cite{zhou}, in the presence
of SOC, Eq. ({\ref{landau}}) is no longer valid. Therefore we calculate the superfluid density by the response of the system with respect to a small phase field of the order parameter: $\Delta \left( \mathbf{r},\tau \right) =\Delta _{0}e^{i\phi}$\cite{fisher}. In \cite{taylor}, E. Taylor proved that this method is equivalent to the definition of superfluid tensor from the current-current correlation function. Furthermore, this method can simultaneously give the compressibility.

By substituting the ansatz $\Delta \left( \mathbf{r},\tau \right) =\Delta _{0}e^{i\phi}$ into the partition function and expanding the partition function to the quadratic order of the phase field $\phi$, after direct but lengthy algebraic manipulations, we obtain the effective action for the phase field as:
\begin{equation}
\mathcal{S}_{eff}\left[ \varphi, \mathbf{A}\right] \simeq \mathcal{S}
_{0}+\int dx\left( \sum_{i=x,y}\frac{\rho_{s}^{i}}{2m}\mathbf{A}_{i}^{2}+\kappa \varphi^{2}\right)
\end{equation}
with the emergent vector field $\mathbf{A}=\mathbf{\nabla}\phi$ and scalar field $\varphi=\nabla \phi$ denoting the spatial and temporal fluctuations of the phase field of the order parameter, respectively. The superfluid tensor can be expressed as: $\rho_{s}^{i}=n - \rho_{n}^{i}$ with the normal density $\rho_{n}^{i}$ are given by:
\begin{eqnarray}
\rho_{n}^{x} &=&\frac{1}{mV}\sum_{\mathbf{k,}s=\pm }k_{x}^{2}Y\left( \mathcal{E}_{\mathbf{
k},s}\right) \frac{\left( \mathbf{M}^{2}+sm\lambda _{x}^{2}\xi _{\mathbf{k}
}\right) ^{2}}{\mathbf{M}^{4}} \\
&+&\frac{m\lambda _{x}^{2}}{2V}\sum_{\mathbf{k,}s=\pm }\tanh \frac{\beta
\mathcal{E}_{\mathbf{k},s}}{2}\frac{\left( \xi _{\mathbf{k}}^{2}+\Delta
_{0}^{2}+s\mathbf{M}^{2}\right) \mathbf{M}_{y}^{4}}{s\mathcal{E}_{\mathbf{k},s}\mathbf{M}^{6}} \label{super_x}
\end{eqnarray}
\begin{eqnarray}
\rho _{n}^{y} &=&\frac{1}{mV}\sum_{\mathbf{k,}s=\pm }k_{y}^{2}Y\left( \mathcal{E}_{\mathbf{
k},s}\right) \frac{\left( \mathbf{M}^{2}+sm\lambda _{y}^{2}\xi _{\mathbf{k}
}\right) ^{2}}{\mathbf{M}^{4}} \\
&+&\frac{m\lambda _{y}^{2}}{2V}\sum_{\mathbf{k,}s=\pm }\tanh \frac{\beta
\mathcal{E}_{\mathbf{k},s}}{2}\frac{\left( \xi _{\mathbf{k}}^{2}+\Delta
_{0}^{2}+s\mathbf{M}^{2}\right) \mathbf{M}_{x}^{4}}{s\mathcal{E}_{\mathbf{k}
,s}\mathbf{M}^{6}} \label{super_y}
\end{eqnarray}
and the compressibility $\kappa$ reads:
\begin{equation}
\kappa =\frac{1}{2V}\sum_{\mathbf{k,}s=\pm }\frac{\xi _{\mathbf{k}}^{2}\left(
\left\vert \Gamma _{\mathbf{k}}\right\vert ^{2}+s\mathbf{M}^{2}\right)
^{2}Y\left( \mathcal{E}_{\mathbf{k},s}\right) }{\mathcal{E}_{\mathbf{k}%
,s}^{2}\mathbf{M}^{4}} +\frac{\Delta _{0}^{2}}{4V}\sum_{\mathbf{k,}s=\pm }\tanh \frac{\beta
\mathcal{E}_{\mathbf{k},s}}{2}\frac{1}{\mathcal{E}_{\mathbf{k},s}^{3}}
\end{equation}
with $\mathbf{M}^{2} =\xi _{\mathbf{k}}\left\vert \Gamma _{\mathbf{k}
}\right\vert $, $\mathbf{M}_{x,y}^{2} =\xi _{\mathbf{k}}\left\vert \Gamma _{\mathbf{k}}^{x,y}\right\vert$, $\left\vert \Gamma _{\mathbf{k}}^{x,y}\right\vert =\lambda _{x,y}k_{x,y}$ and $Y\left(x\right)=\beta f(x)[1-f(x)]$.

We first self-consistently solve Eq. ({\ref{gap}}) and Eq. ({\ref{num}}) to get the chemical potential and gap parameters and then substitute the obtained results into Eq. ({\ref{gap}}) and Eq. ({\ref{num}}) to get the condensed fraction and superfluid fraction tensor.

\section{Result and discussions}
In this paper, we only consider the ground state properties of the system. Previous investigations show that mean-field theory is qualitatively and quantitatively correct in the low temperature regime. However, at finite temperatures, fluctuations of the order parameters become more and more important. In order to get qualitatively correct physics for temperature close to critical temperature, the most successful method is to include contributions in the lowest gaussian fluctuations of the gap parameter to the thermodynamic potential which is beyond the scope of this paper and will be left as a future work.

\begin{figure}[htp]
\includegraphics[width=\columnwidth,height=50mm]{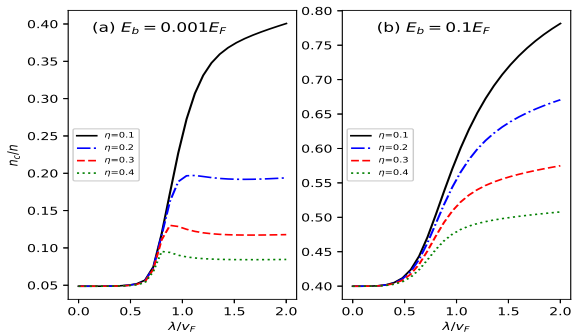}
\caption{(Color online) Condensed fraction defined by: $n _{c}/n$ as functions of SOC strength parameter $\lambda/v_F$ with $v_F$ being the Fermi velocity.}
\label{nc_lam}
\end{figure}

\begin{figure}[htp]
\includegraphics[width=\columnwidth,height=50mm]{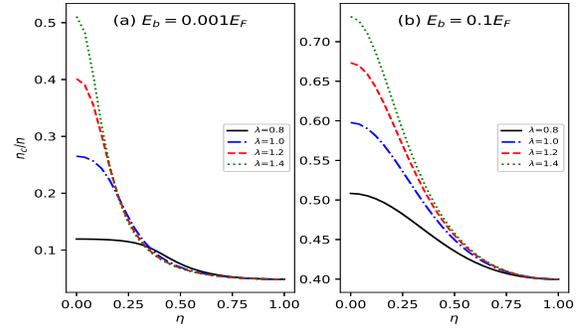}
\caption{(Color online) Condensed fraction as functions of anisotropic parameter $\eta=\lambda_D/\lambda_R$.}
\label{nc_eta}
\end{figure}

In Fig. \ref{nc_lam}, we present our numerical results of the condensed fraction as a function of the SOC parameter $\lambda/v_{F}$ with $v_{F}=p_{F}/m$ being the Fermi velocity and $p_{F}$ being the Fermi momentum. It is clear that SOC enhances condensation compared with cases with no SOC. However, the condensed fraction shows non-monotonic behaviors for some parameters space. In Fig. \ref{nc_lam}(a) , the interaction parameter is $E_b = 0.001E_{F}$ where $E_F=p_F^2/2m$ is the Fermi energy. As one can see that in this weak interaction regime, as we increase the anisotropic parameters, the condensed fraction decreases but has a maximum value as functions of $\lambda/v_{F}$. This means that for large enough anisotropic parameters, SOC does not necessarily enhances condensation. However, in the strong interaction limit with large enough $E_{b}$ shown in Fig. \ref{nc_lam}(b), condensed fraction is always a monotonic function of SOC strength. Fig. \ref{nc_eta} represents condensed fraction as functions of anisotropic parameters. And one can see that anisotropic parameters always suppresses condensation. In Fig. \ref{nc_eta}(a), different lines crosses with each other, which is a direct manifestation of the fact that for large enough anisotropic parameters, condensed fraction is not a monotonic function with respect to SOC strength.

In general, as a static property, the condensed fraction has similar behaviors as the gap parameters.  However, the superfluid fraction tensor becomes more complicated for the superfluid Fermi systems with anisotropic SOC. It is easily seen from Eq. \ref{super_x} and Eq. \ref{super_y} that SOC suppresses superfluidity and it creates normal density even at zero temperature.

\begin{figure}[htp]
\includegraphics[width=\columnwidth,height=60mm]{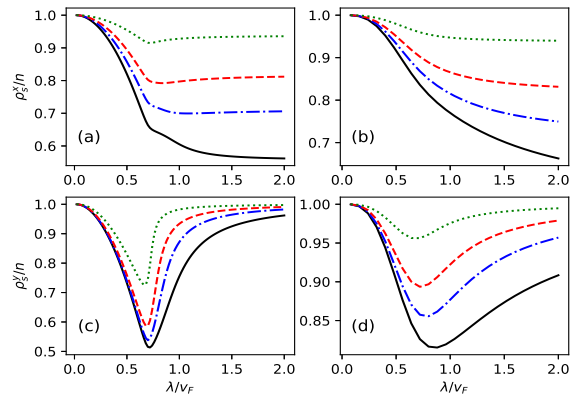}
\caption{(Color online) Superfluid fraction $\rho_s^{x,y}/n$ as functions of the SOC strength parameter $\lambda/v_F$. Real black lines, dot-dashed blue lines, dashed red lines and dotted green lines correspond to $\eta=0.2$, $\eta=0.3$, $\eta=0.4$ and $\eta=0.6$, respectively. $E_b=0.001E_F$ in (a) and (c). $E_b=0.1E_F$ in (b) and (d).}
\label{superxy_lam}
\end{figure}

Fig. \ref{superxy_lam} represents numerical results for the superfluid fraction tensor $\rho_{s}^{x}$ and $\rho_{s}^{y}$ as functions of the SOC strength for various interaction and anisotropic parameters. We can see from Fig. \ref{superxy_lam} (a) with $E_b=0.001E_F$ that $\rho_{s}^{x}$ decreases with increasing SOC strength for small anisotropic parameters in the weak interaction limit. However, for large anisotropic parameters, $\rho_{s}^{x}$ is a non-monotonic function of SOC strength with a global minimum. And this minimum for large anisotropic parameters disappears for strong interaction parameters as shown in Fig. \ref{superxy_lam} (b) with $E_b=0.1E_F$. Nonetheless, the other component of the superfluid tensor $\rho_{s}^{y}$ has different behaviors as shown in Fig. \ref{superxy_lam} (c) and (d). We have checked for various parameters and found that $\rho_{s}^{y}$ always has a minimum regardless of the value of the anisotropic and interaction parameters. It is also clear that $\rho_{s}^{x,y} \rightarrow n$ for $\lambda = 0$.

\begin{figure}[htp]
\includegraphics[width=\columnwidth,height=50mm]{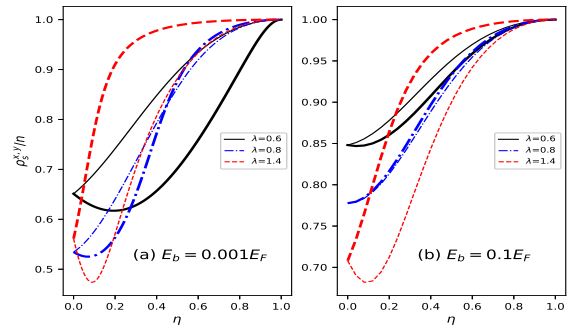}
\caption{(Color online) Superfluid fraction as functions of anisotropic parameters. In both figures, thick lines and thin lines correspond to $\rho_s^{y}/n$ and $\rho_s^{x}/n$, respectively.}
\label{superxy_eta}
\end{figure}

We also investigated effect of anisotropy on the superfluid fraction and the results are shown in Fig. \ref{superxy_eta}. As can be seen clearly from the results, $\rho_{s}^{x}$ and $\rho_{s}^{y}$ as functions of anisotropic parameters are more complicated. Firstly, in Fig. \ref{superxy_eta} (a) for weak interaction parameter $E_b=0.001E_F$, $\rho_{s}^{x}>\rho_{s}^{y}$ and only $\rho_{s}^{y}$ shows a minimum form small SOC parameters. For large SOC parameters, $\rho_{s}^{x}<\rho_{s}^{y}$ and only $\rho_{s}^{x}$ has a minimum. Secondly, when the system enters strong interaction regime, as shown in Fig. \ref{superxy_eta} (b) with $E_b=0.1E_F$, $\rho_{s}^{x}>\rho_{s}^{y}$ but with no minimum for $\rho_{s}^{y}$ for weak SOC. And for large SOC parameters, $\rho_{s}^{x}<\rho_{s}^{y}$ and $\rho_{s}^{x}$ still shows a minimum. And we have checked for larger value of interaction parameters ($E_b=1.0E_F$), the situations are the same as shown in Fig. \ref{superxy_eta} (b). $\rho_{s}^{x}$ always has a minimum value for large SOC parameters. Finally, we noticed that $\rho_{s}^{x}=\rho_{s}^{y}$ with $\eta =0$ and we have the isotropic Rashba SOC case where $\rho_{s}^{x}=\rho_{s}^{x}$. Furthermore, $\rho_{s}^{x}=\rho_{s}^{y}=1$ for the anisotropic case with $\eta=1$. This is true since in the ERD case, the SOC hamiltonian density reduces to a one-dimensional SOC term. For balanced case with equal particle numbers for spin down and spin up atoms, this one-dimensional SOC term can be gauged out and has no effect on the thermodynamic properties of the system. Therefore, the superfluid fraction at zero temperature goes to $1$ as in the case without SOC.

A final remark: as shown in the numerical results of the superfluid fraction tensor, $\rho_{s}^{x}$ and $\rho_{s}^{y}$ have different behaviors. This comes from the fact that we constraint the anisotropic parameter in the domain: $0<\eta<1$. Therefore, we never reach the regime for pure Dresselhaus limit. In the transparent anisotropic expression for the SOC part of the Hamiltonian density, $0<\eta<1$ means $\lambda_R > \lambda_D$ and $\lambda_y > \lambda_x$. The other limit of pure Dresselhaus SOC can be reached by setting $\eta \rightarrow \infty$. For symmetry considerations, thermodynamic properties should be symmetric about the two regime: $0<\eta<1$ and $1<\eta<\infty$.

\section{Conclusion and outlook.}

We performed a detailed research on the effectf of an anisotropic SOC on the condensation and superfluid properties of a two dimensional
Fermi gases at zero temperature. Particularly, we found that
SOC not always enhances condensation and suppresses superfluidity. The condensed fraction and superfluid tensor show many different behaviors for different parameter configurations. In this paper, we only consider the phase fluctuations of the order parameter and neglect the magnitude fluctuations. Besides, inclusion of optical lattice would give us much degrees of freedom and more interesting phenomena such as the superfluid-Mott insulator transition. Furthermore, if we consider imbalanced case, there will be topological phase transition as we increase the Zeeman field across a critical value. Combinations of SOC and optical lattice provides a ideal test ground for many interesting phenomena observed in solid state material system.

\section{Acknowledgements.}
This work has been supported by the Scientific Research Foundation of Hunan Provincial Education Department under Grant number 20C0648.
\bibliographystyle{ieeetr}
\bibliography{reference}
\end{document}